\documentclass{osa-article}

\journal{osac}

\articletype{Research Article}

\begin{document}

\title{An Optical Voltage Sensor Based on Piezoelectric Thin Film for Grid Applications}

\author{Jordan L. Edmunds,\authormark{1,*} Soner Sonmezoglu,\authormark{1,**} Julien Martens,\authormark{2} Alexandra Von Meier,\authormark{1} Michel M. Maharbiz,\authormark{1,3,4}}

\address{\authormark{1}Department of Electrical Engineering and Computer Sciences, University of California, Berkeley, Berkeley, CA 94704, USA\\
\authormark{2}Department of Microsystems Engineering (IMTEK), Laboratory for Biomedical Microtechnology, University of Freiburg, Freiburg, Germany\\
\authormark{3}Department of Bioengineering, University of California, Berkeley, Berkeley, CA 94704, USA\\
\authormark{4}Chan Zuckerberg Biohub, San Francisco, CA 94158, USA\\}

\email{\authormark{*}jordan.e@berkeley.edu} 
\email{\authormark{**}ssonmezoglu@berkeley.edu} 



\begin{abstract}
Continuous monitoring of voltages ranging from tens to hundreds of kV over environmental conditions, such as temperature, is of great interest in power grid applications. This is typically done via instrument transformers. These transformers, although accurate and robust to environmental conditions, are bulky and expensive, limiting their use in microgrids and distributed sensing applications. Here, we present a millimeter-sized optical voltage sensor based on piezoelectric aluminum nitride (AlN) thin film  for continuous measurements of AC voltages <350$kV_{rms}$ (via capacitive division) that avoids the drawbacks of existing voltage-sensing transformers. This sensor operated with 110$\mu W$ incident optical power from a low-cost LED achieved a resolution of 170$mV_{rms}$ in a 5kHz bandwidth, a measurement inaccuracy of 0.04\% due to sensor nonlinearity, and a gain deviation of +/-0.2\% over the temperature range of \textasciitilde 20-60$^{\circ}C$. The sensor has a breakdown voltage of 100V, and its lifetime can meet or exceed that of instrument transformers when operated at voltages <$42V_{rms}$. We believe that our sensor has the potential to reduce the cost of grid monitoring, providing a path towards more distributed sensing and control of the grid.
\end{abstract}

\section{Introduction}

Safe, accurate, and economical measurement of time-varying voltages in electric power systems is a significant challenge. The standard solution to this challenge is the instrument transformer, which steps down a high voltage (\textasciitilde 1kV+) to an appropriate level (typically <100V) and isolates the stepped-down voltage, allowing safe measurement using conventional electronics. However, these transformers are bulky and expensive and sometimes explode (\textasciitilde 3\% of all installed instrument transformers  \cite{poljak2010method}). Optical methods for direct measurement of high voltages have gained  attention since the early  1980s \cite{yoshino1982fiber}, mainly due to the high available bandwidth (\textasciitilde GHz), intrinsic electrical isolation, and the potential for low cost and remote monitoring. Initial optical voltage sensors consisted of long (several meters) optical fiber wrapped around a piezoelectric material \cite{yoshino1982fiber}. In these sensors, when the piezo material was excited by a voltage, it grew or shrank proportional to the applied voltage, changing the fiber optical path length; the resulting change in the optical path length was measured using interferometry to infer the voltage amplitude. More recent optical methods include coupling piezoelectric material to the resonant frequency of Bragg fiber gratings \cite{seeley2007packaging, yang2017optical, gonccalves2019temperature}. However, the sensors based on the above optical methods are inaccurate due to their nonlinearities \cite{gonccalves2019temperature} or temperature sensitivity \cite{yang2017optical, gonccalves2019temperature, dante2016temperature}. Closed-loop compensation methods were used to improve the accuracy of such sensors \cite{hui2013tracking, li2014analysis}, but these methods reduce system reliability and increase system complexity, electrical hazards, and cost.

Fundamentally, output nonlinearity and temperature dependence increase with an increasing quality factor ($Q$) of the interferometric- or optical cavity-based sensor, where Q is defined as the ratio of the energy stored in the cavity to the energy dissipated per oscillation period. Therefore, a low-$Q$ sensor is desirable to minimize nonlinearity and temperature dependence. However, low-$Q$ sensors suffer from low sensitivity, resulting in a low signal-to-noise ratio (SNR) and poor performance. In practice, the SNR, and hence performance, can be improved independently of the sensor sensitivity by increasing the incident optical power ($P_{in}$) or reducing the operating bandwidth ($BW$). This makes performance comparison of different optical voltage sensors difficult; they must be operated with the same $P_{in}$ and $BW$ for a fair comparison.

Given this, a $P_{in}$ and $BW$ independent figure of merit (FoM) is required to quantify the noise performance of optical voltage sensors. We propose such a metric, the \textit{energy per quanta} ($E_Q$), which depends only on the sensor properties and dynamic input range. This metric extends the power efficiency factor used in analog circuit design \cite{muller20110} and the energy per conversion-level of analog-to-digital-converters \cite{walden1994analog, walden1999analog}. Lower $Q$-factors yield less sensitive sensors and a higher $E_Q$ ($E_Q \propto 1 / Q^2 $).

In this work, we propose trading off $E_Q$ for temperature insensitivity and reduced harmonic distortion, and explore the limits of this approach. To this end, we demonstrate a low-$Q$ resonant optical cavity-based voltage sensor based on a piezoelectric AlN thin film that transduces a voltage applied across the piezo terminals into a change in the resonant frequency of the cavity. This sensor can be batch fabricated with high yield and low cost (\textless\$1), which makes it uniquely well-suited to reduce the cost of grid voltage measurement.

%

\section{Optical voltage sensor design and sensor fabrication}
\subsection{Operating principle and fabrication process of the sensor}

\begin{figure}[h]
\begin{center}
  \includegraphics[width=4.5in]{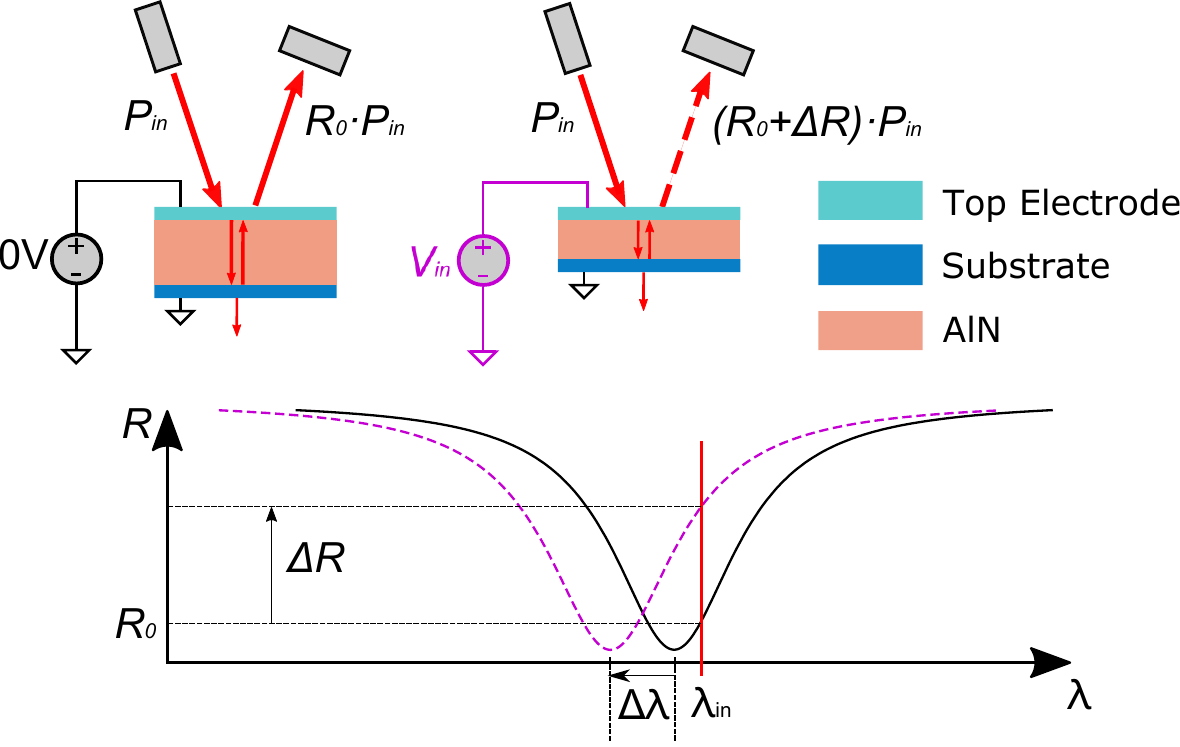}
  \caption{ Sensor principle of operation. During operation, the sensor is illuminated by an incident optical power $P_{in}$ at a wavelength $\lambda_{in}$ (that is, the steepest point of the reflectance curve) offset from the resonant wavelength $\lambda_{r}$. An input voltage $V_{in}$ applied across the cavity results in a shift in the resonant spectrum from the solid (unperturbed) to the dashed (perturbed) spectrum, causing a change in the reflectance $\Delta R$ at $\lambda_{in}$. The resulting $\Delta R$ change alters the amount of reflected light power ($P_{r}$=$R_{0}$+$\Delta R) \cdot P_{in}$) that is measured to determine the $V_{in}$. }
  \label{fig:ovs-principle-of-operation}
\end{center}
\end{figure}

Fig. 1 shows the operating principle of the proposed optical voltage sensor (OVS) based on changes in the measured reflectance of a resonant cavity, whose thickness varies with applied voltage. The resonant cavity is formed by an AlN thin film sandwiched between the top indium tin oxide (ITO) electrode and the bottom silicon (Si) substrate. During operation, the sensor is illuminated by a light source with an incident optical power ($P_{in}$) at a fixed wavelength ($\lambda_{in}$) near the resonance wavelength of the cavity ($\lambda_{r}$). Some fraction of $P_{in}$ is reflected from the cavity, with the remainder dissipated in or transmitted through the cavity, as seen in Fig. \ref{fig:ovs-principle-of-operation}. Here, the intensity of reflected light ($P_{r}$=$P_{in}$×$R$, where $R$ is the cavity reflectance) is measured by a photodetector to detect the amplitude of an input voltage ($V_{in}$) applied across the cavity. $P_{r}$ depends on $V_{in}$ through $R$ as $V_{in}$ generates an electric field in the cavity that changes the AlN film thickness \cite{lueng2000piezoelectric} and refractive index  \cite{graupner1992electro} and hence results in a shift in resonant wavelength $\Delta \lambda = \lambda_{r} - \lambda_{r0}$, where $\lambda_{r0}$ is the resonant wavelength at $V_{in}=0$. The resulting $\lambda_{r}$ shift leads to a change in the reflectance ($\Delta{R}$ =$R$-$R_0$, where $R_0$ is the reflectance at $V_{in}$=0). The $\Delta{R}$ value at a known $V_{in}$ can be calculated using the following expression (see supplementary material for the derivation of (1)): 

\begin{equation}
    \Delta R = \beta V_{in}
\end{equation}

\begin{equation}
    \label{eq:beta-equation}
    \beta = \frac{3 \sqrt{3}}{4} \frac{R_{max} Q}{t} \left(d_{33} + \frac{1}{2} n_{0}^2 r_{33} \right)
\end{equation}


\noindent where $R_{max}$ is the amplitude of the resonant dip (always <1), $Q$ and $t$ are the quality factor and thickness of the cavity, $d_{33}$ is the thickness mode piezoelectric strain coefficient, $n_{0}$ is the (unperturbed) refractive index of the AlN thin film, and $r_{33}$ is the Pockels coefficient (which relates the refractive index to the applied electric field). This equation is valid near $\lambda_{in} \approx \pm \lambda_r + 1/\sqrt{3} \cdot FWHM$; this corresponds to the steepest point on the reflectance curve, where $FWHM$ is the full-width-half max of the resonant dip.

\begin{figure}[h]
\begin{center}
  \includegraphics[width=5.0in]{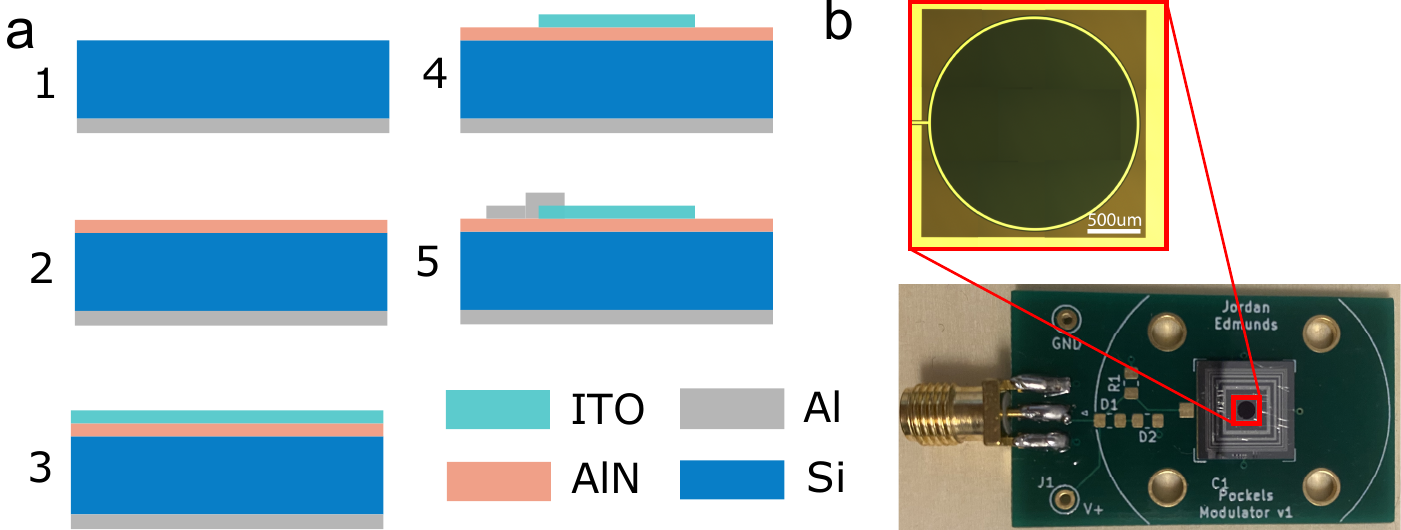}
  \caption{ (a) Sensor fabrication process. 1. Backside Al (\textasciitilde 300nm) sputter and anneal at 300°C for 15 minutes 2. Frontside AlN (300nm) reactive sputter 3. Topside ITO (20nm) sputter 4. Lithographic patterning, ion milling of device mesas 5. Bond pad patterning Ti/Al (20nm/300nm) evaporation and lift-off. (b) Photograph of the fabricated sensor die on a PCB, and (inset) optical top-view micrograph of the sensor. }
  \label{fig:ovs-device-fabrication}
\end{center}
\end{figure}

Fig. \ref{fig:ovs-device-fabrication}(a) shows the OVS fabrication process. All lithography was done in a DUV stepper (ASML), on a 150mm-thick silicon (Si) wafer. We first sputtered a \textasciitilde 300nm-thick layer of backside aluminum (Al), and then annealed the wafer at 300°C for 15min in atmosphere to create backside ohmic contacts that serves as the bottom electrode. Next, we deposited a 300nm-thick aluminum nitride (AlN) film (endeavor AT) and a 20nm-thick indium tin oxide (ITO) film to serve as the transparent top electrode. Finally, we patterned the top ITO contacts to form 2mm diameter devices and evaporated 20nm titanium (Ti)/300nm Al to form bond pads. The fabricated 10×10$mm^2$ sensor die was attached with conductive silver epoxy and wire-bonded to a printed circuit board (PCB). An optical micrograph of the 2mm diameter sensor on its PCB are shown in Fig. \ref{fig:ovs-device-fabrication}(b) The die size was designed to be larger than the actual device size to facilitate easy handling. See the supplementary information for specific process details.

\textbf{Optical shot noise sets a hard limit on OVS performance.} The performance of optical detection systems is bounded by shot noise received at the photodetector. For a shot-noise limited OVS system, the system SNR is proportional to the input optical power, which is given by $I_{pd}^2/I_{noise}^2 = \Delta I_{pd}^2/(2q (I_{pd}+ \Delta I_{pd}) \cdot BW)$, where $q$ is the electron charge, $I_{pd}$ is the light-induced photocurrent on the photodetector, $\Delta I_{pd}$ is the change in the rms photocurrent induced by an applied input voltage ($V_{in}$), and $BW$ is the system bandwidth. The SNR can also be expressed in terms of an an incident light power ($P_{in}$), the average device reflectance ($ \langle R \rangle$), and an rms modulation depth ($\Delta R$, that is, the change in $R$ due to an applied $V_{in}$) in the following form, using $I_{pd} = P_{in} \cdot \Re \cdot R$:


\begin{equation}
SNR \approx \frac{P_{in} \cdot \Re \cdot \Delta R^2}{2q \cdot \langle R \rangle \cdot BW}
\label{eq:snr-equation}
\end{equation}

\noindent where $\Re$ is the responsivity of the photodetector. Eq. \ref{eq:snr-equation} reveals the system SNR depends not only on $V_{in}$ through $R$ but also $P_{in}$ and $BW$, consistent with the expectation that larger $P_{in}$ and smaller $BW$ provide a better SNR in the optical voltage sensing system. However, optical sources can only supply a limited amount of power, and system-level requirements could potentially limit the maximum $P_{in}$ and the minimum $BW$ in the system.

\textbf{The input-referred energy per quanta ($E_Q$) allows quantitative configuration-independent sensor comparison.} Since the best choices for $P_{in}$ and $BW$ will vary by application, it is useful to introduce a metric that normalizes SNR to $P_{in}$ and $BW$. This would allow a rigorous noise performance comparison between optical voltage sensors, independent of the sensor's particular operating $P_{in}$ or $BW$. In digital systems (optical and otherwise), the energy per bit has become a ubiquitous metric of device performance \cite{tucker2010green}. Here, we propose an alternative metric for analog systems, the energy per quanta ($E_Q$), defined as:

\begin{equation}
\label{eq:Esdef}
    E_{Q} \equiv  \frac{P_{in}}{\mathrm{SNR} \cdot BW}
\end{equation}

This metric demonstrates how efficiently $P_{in}$ is used, and can be interpreted as a cost paid in energy to achieve a desirable SNR. For a shot-noise limited system, $E_Q$ can be derived, independent of $P_{in}$ and $BW$, by inserting Eq. \ref{eq:snr-equation} into Eq. \ref{eq:Esdef}:

\begin{equation}
\label{eq:Esmod}
    E_{Q,min} \approx \frac{q \cdot \langle R \rangle}{\Re \cdot \Delta R^2}
\end{equation}

The form of this equation makes it clear that reducing the average cavity reflectance $\langle R \rangle$ at the expense of modulation depth $\Delta R$ at the operating wavelength can improve the noise performance of the system, as observed in \cite{lee2009optimization}. $E_Q$ is bounded from below by the incident photon energy captured by the photodetector.

\subsection{Optical voltage sensor (OVS) design}
We designed our OVS to measure grid-level AC voltages in the range of tens to hundreds of kVs via capacitive division. The system bandwidth ($BW$) was set to 5kHz, satisfying the $BW$ requirement of most grid applications, including inverter-based solar \cite{fan2020subcycle}. To minimize sensor nonlinearity and temperature sensitivity, we chose to deliberately design the sensor to have a high $E_Q$, as $d\hat{\beta}/dT \propto 1 / E_Q$ and $d\hat{\beta}/dV_{in} \propto 1 / E_Q$ (where $\beta$ is the sensor gain, shown in Eq. 2, and $\hat{\beta} = \beta(V_{in},T) / \beta(0,0)$ is the normalized gain). To achieve a high $E_Q$, we minimized the sensor $Q$-factor by designing the cavity as thin as possible ($L=300nm$) and excluding mirrors (other than the material interfaces).

Furthermore, most previous sensors use lead zirconate titanate (PZT) as a piezoelectric material to form the resonant cavity as it provides a large piezoelectric strain coefficient ($d_{33}\approx500pm/V$ \cite{du1998crystal}). However, the PZT $d_{33}$ is extremely temperature sensitive (\textasciitilde20\% over 100$^{\circ} C$ ) \cite{li2009determination}. Therefore, here we elected to use aluminum nitride (AlN) for our sensor to further minimize the sensor temperature dependence as its $d_{33}$ is independent of temperature \cite{kano2006temperature, rossel2014temperature}. 


\section{Experiment results}

\begin{figure}[h]
\begin{center}
  \includegraphics[width=5in]{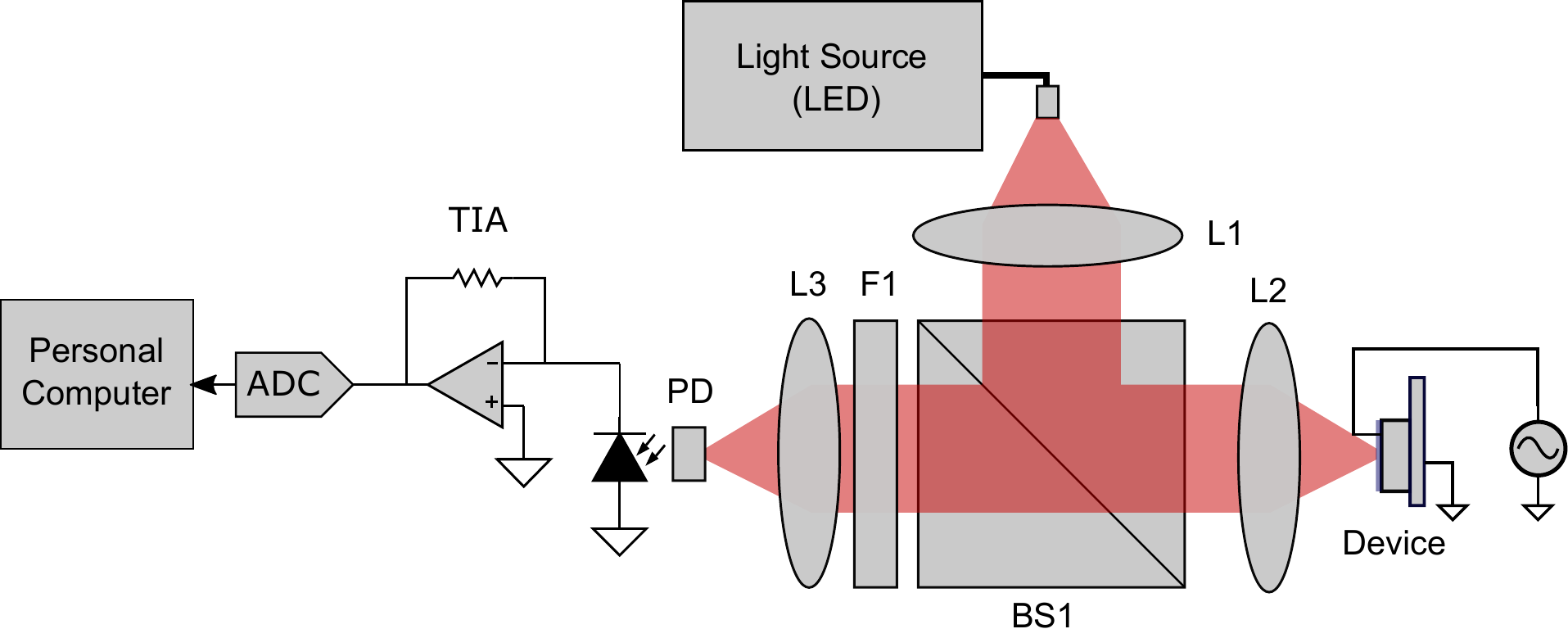}
  \caption{ Sensor characterization setup. Light is emitted by an LED source, focused onto device, imaged onto a photodiode, and measured with a TIA. The TIA output is sent to the ADC to be recorded in the computer. }
  \label{fig:ovs-device-characterization}
  \end{center}
\end{figure}

The optical voltage sensor (OVS) was characterized using the setup shown in Fig. \ref{fig:ovs-device-characterization}, where we used an LED (Thorlabs, M970F3) with a peak intensity at \textasciitilde$970nm$ ($\pm 10nm$) and a mean intensity at \textasciitilde 950nm. The light from the LED is fiber-coupled to collimating lens L1, beam-splitter BS1, focusing lens L2, focusing lens L3, 850nm long-pass filter F1 (Thorlabs FELH0850) and Si photodiode PD (Thorlabs, SM05PD2B). The PD (Thorlabs, SM05PD2B) was connected to a transimpedance amplifier (TIA) with a 1M$\Omega$ feedback resistance to convert a light-induced photocurrent on the PD to a voltage. The resulting voltage was digitized by an analog-to-digital converter (ADC) (NI myDAQ; National Instruments) with a 10kHz sampling rate and then sent to a computer through a serial link for data storage and further analysis.

\begin{figure}
\begin{center}
  \includegraphics[width=5.5in]{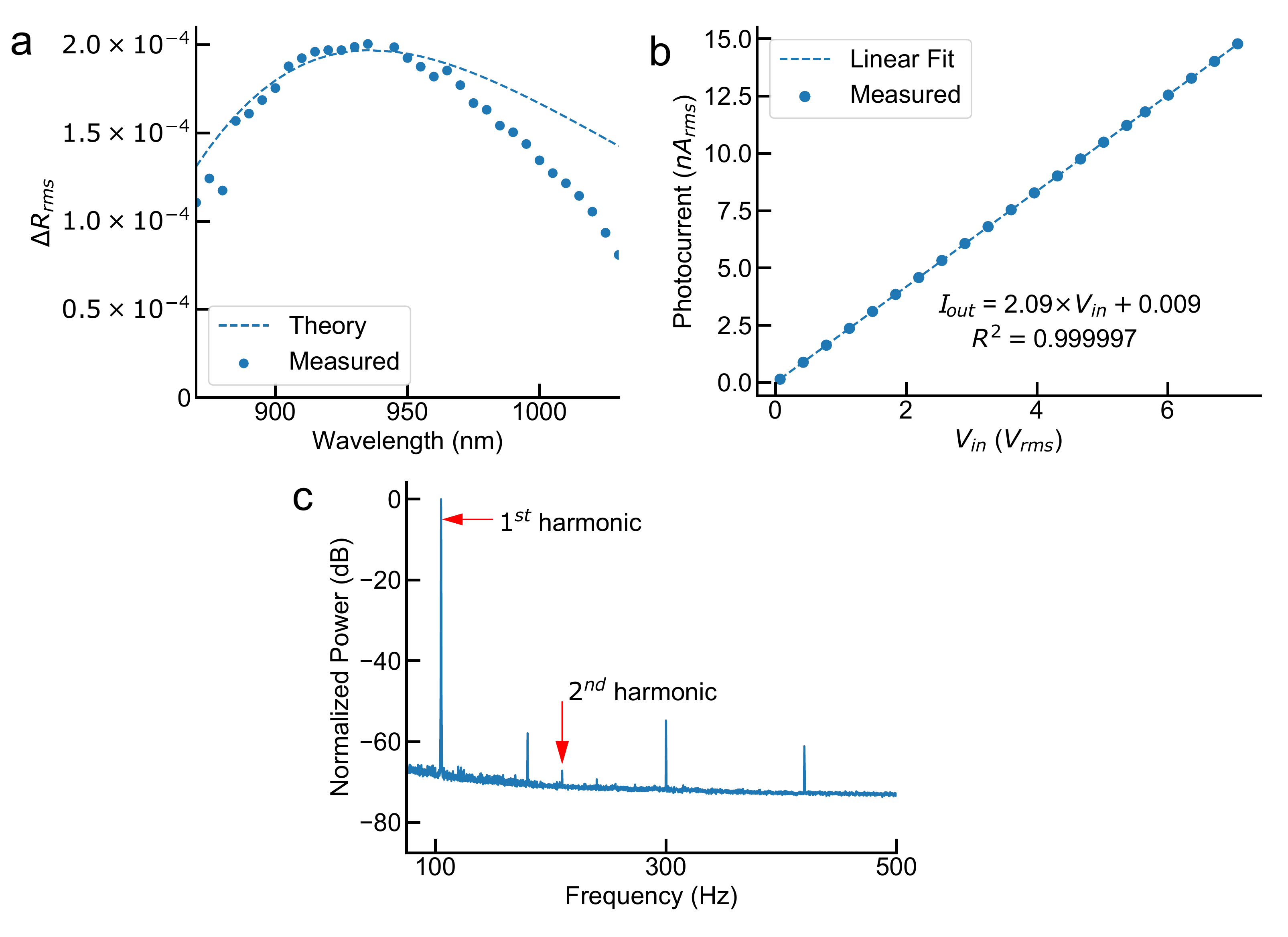}
  \caption{ (a) Measured and predicted modulation depth spectra of the sensor operated with an input voltage ($V_{in}$) of $20V_{pp}$ (Dataset 1). (b) Output photocurrent versus $V_{in}$ applied to the sensor (Dataset 2). (c) Power spectral density of the sensor output (averaged 500 times, the duration of each measurement was 10s), normalized with respect to the first harmonic amplitude (Dataset 3). }
  \label{fig:ovs-modulation-spectra}
\end{center}
\end{figure}

To extract the modulation depth ($\Delta{R_{rms}}$) spectrum of the OVS, we applied a 105Hz sinusoidal signal ($V_{in}$) to the sensor and used a 2.4nm bandwidth monochromator (Dynasil, DMC1-05G) with the LED. Note that the monochromator is only used in the $\Delta{R_{rms}}$ measurement. The collected photocurrent spectrum data was normalized to the data obtained in the same manner using a 120nm gold-coated sample. Fig. \ref{fig:ovs-modulation-spectra}(a) depicts the measured $\Delta{R_{rms}}$ spectra for the sensor operated at $V_{in}$ of $20V_{pp}$, showing good agreement with the $\Delta{R_{rms}}$ spectra predicted from a thin-film Fresnel equation model (supplementary materials) using the parameters provided in Supplementary Table S1. The difference in the measured and predicted spectra can be mainly attributed to the Si substrate becoming increasingly transparent at longer wavelengths, causing the transmitted light through the AlN layer (subsequently reflected by the Si/Al interface) to partially cancel the reflected light from the cavity.


\textbf{The sensor operated with an incident optical power ($\mathbf{P_{in}}$) of $\mathbf{110 \mu W}$ exhibits a resolution of $\mathbf{170mV_{rms}}$ in a 5kHz bandwidth with a full scale of $140V_{rms}$.} Fig. \ref{fig:ovs-modulation-spectra}(b) depicts the output photocurrent as a function of $V_{in}$ (105Hz sinusoidal signal) (Dataset 2). The sensor operated with $P_{in}$ of $110 \mu W$ shows a sensitivity of 2.09nA/V and a noise floor of $5.0pA/\sqrt{Hz}$, yielding a voltage resolution of $2.4mV / \sqrt{Hz}$, corresponding to $170mV_{rms}$ in a 5kHz bandwidth. Note that the measured noise floor was in excess of the shot noise limit ($2.3pA/\sqrt{Hz}$) by 6.6dB, dominated by the noise from the LED.

\textbf{The sensor nonlinearity results in a measurement inaccuracy of only 0.04\%.} In the nonlinearity test, the sensor was operated with $V_{in}$=$20V_{pp}$ and $P_{in}$=\textasciitilde $110 \mu W$. Fig. \ref{fig:ovs-modulation-spectra}(c) shows the normalized power spectral density of the sensor output (Dataset 3). The second harmonic power (-67.3dB) corresponds to a measurement inaccuracy of 0.04\% in the sensor output; this result is consistent with the nonlinearity of AlN reported in \cite{boales2018measurement}. The third harmonic is invisible in the spectrum as its power is below the sensor noise level. The other tones seen in Fig. \ref{fig:ovs-modulation-spectra}(c) are the 60Hz interference tone and its numerous harmonics.


\begin{figure}
  \begin{center}
  \includegraphics[width=3.5in]{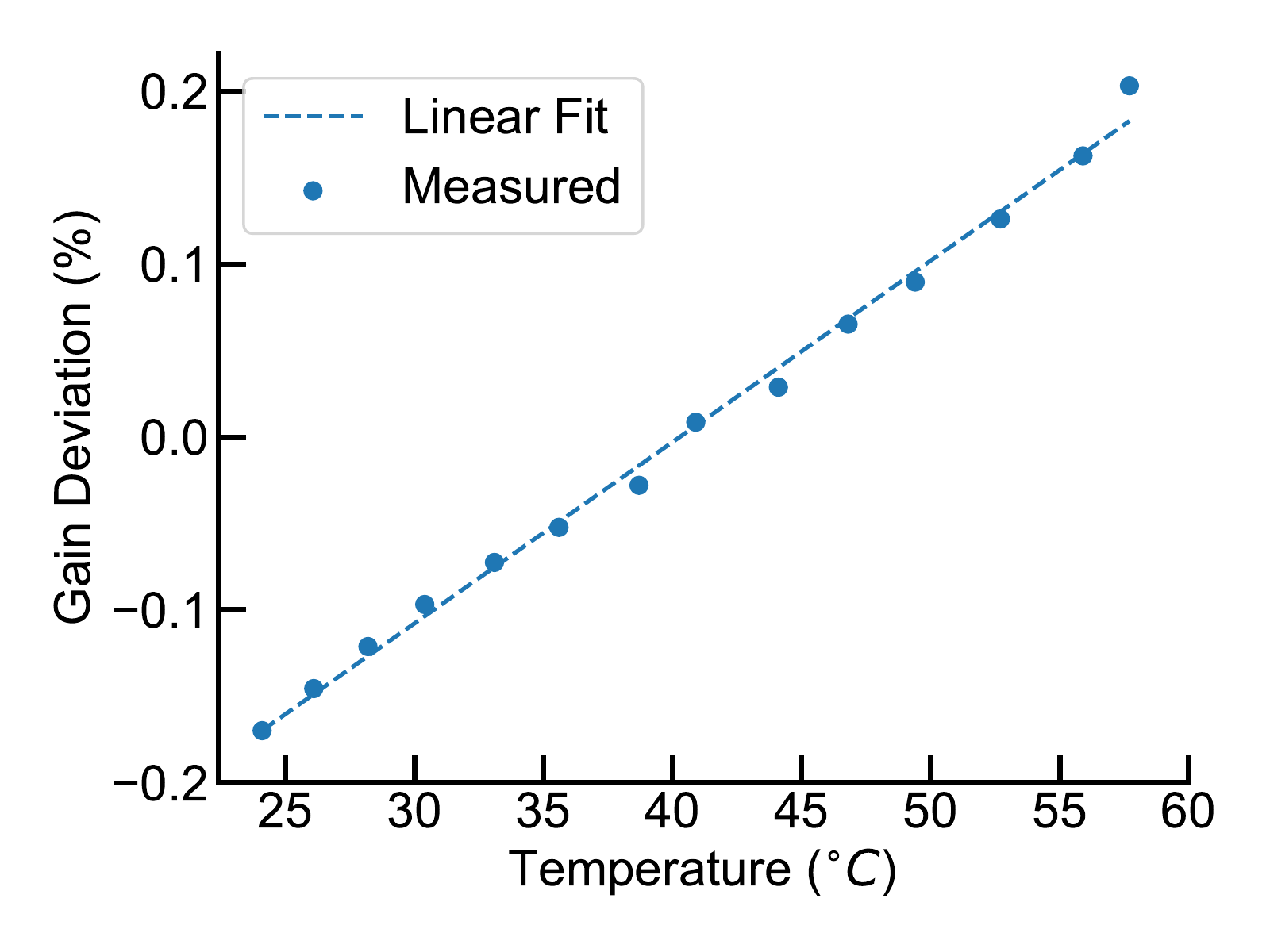}
  \caption{ Normalized change in sensor gain ($\Delta \beta / \beta$) versus temperature. Raw data available in Dataset 4.}
  \label{fig:ovs-temperature-gain-offset}
  \end{center}
\end{figure}

 \textbf{The sensor output varies only +/-0.2\% over a \textasciitilde 40°C temperature range.} We measured the temperature dependence of the sensor gain ($ \beta $, see Eq. 2). During the measurement, the sensor, operated with a $20V_{pp}$, 105Hz sinusoidal signal ($V_{in}$), was placed on a 250µm-thick polyimide heater (Omega) that was controlled using a PID controller (Omega, CSi32) and a K-type thermocouple attached to the sensor die. The measurement result in Fig, \ref{fig:ovs-temperature-gain-offset} showed that the sensor gain varies only +/-0.2\% in the temperature range of 23-57$^{\circ}C$, yielding a temperature sensitivity of \textasciitilde $0.01\%/^{\circ}C$. We fit the measured $ (\Delta \beta / \beta) / \Delta T $ using a monochromatic Fresnel equation based optical model \cite{hecht2002optics}, with the incident wavelength as the fitting parameter. From this model, the incident wavelength was fit to 940nm, which is 10nm lower than the LED's mean emission wavelength, within the manufacturer's tolerance. This is about 30nm away from the optimal operation wavelength of 911nm at which point the error is predicted to be quadratic with temperature, with a max deviation of less than +/-0.02\%.

\begin{figure}
\begin{center}
  \includegraphics[width=3.5in]{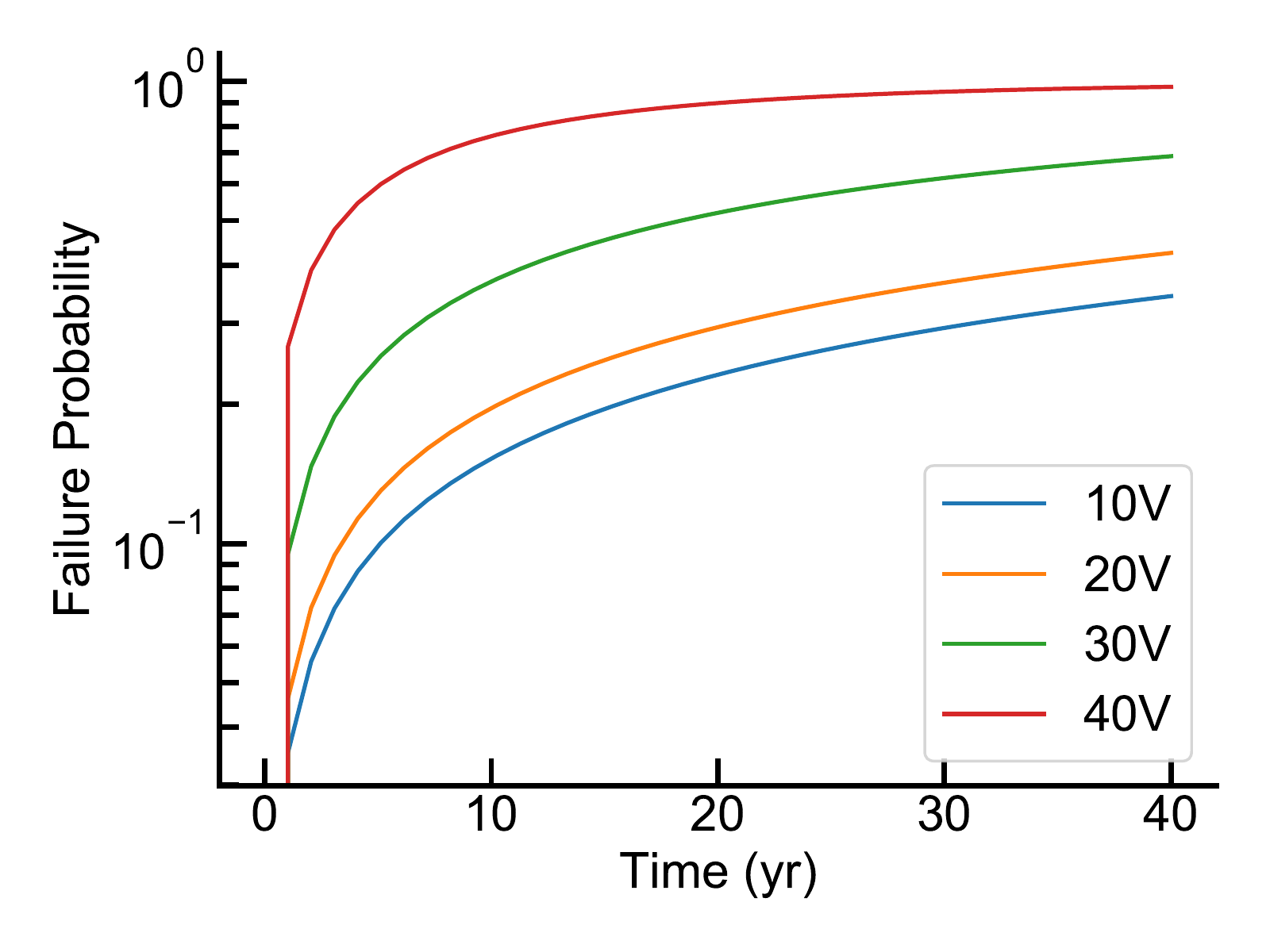}
  \caption{ Probability of device failure versus time at various DC operating voltages. Data underlying curve available in Datasets 6-7}
  \label{fig:ovs-device-lifetime}
\end{center}
\end{figure}

\textbf{Sensor lifetime can meet or exceed that of instrument transformers.} To determine the maximum electric field ($E$) that could be safely applied to the sensor along with the expected sensor lifetime, we measured the breakdown charge $Q_{bd}$ of 8 sensors with low leakage currents of less than 0.1nA at 10V input voltage ($V_{in}$) \cite{verweij1996dielectric}. During the measurement, we subjected these sensors to a linear voltage ramp to $43V$, followed by a temperature ramp (\textasciitilde$5^{\circ}C/min$) to $180^{\circ}C$, and recorded the current over time until the point of failure. The obtained data were fit to a Weibull distribution \cite{mccool2012using}, whose cumulative distribution function (CDF) is

\begin{equation}
1 - e^{{-\left(Q/Q_0\right)}^\gamma}
\end{equation}

\noindent where $\gamma$ is the Weibull slope and $Q_0$ is the characteristic charge. We extracted values of $\gamma$=0.67 and $Q_0$=379mC (Supplementary figure S2). Combining the extracted Weibull CDF (Supplementary figure S2) with the leakage current at room temperature (Supplementary figure S3), we estimated the likelihood of sensor failure over time at various operating input voltages (see Fig. \ref{fig:ovs-device-lifetime}).

\section{Discussion}

\textbf{Monitoring grid-level voltage up to \textasciitilde 350$kV_{rms}$ is possible through capacitive division.} Since the optical voltage sensor (OVS) is based on a 300nm AlN thin film, a large device capacitance (>1nF) is possible despite the small sensor size (2mm diameter) and relatively low-permittivity dielectric ($\epsilon_r$=9.5). We measured the sensor impedance using an HP2484A LCR meter from 10kHz-500kHz (Supplementary figure S4, Dataset 5), and extracted the device capacitance (C=0.997nF) and resistance (R=389$\Omega$) from fitting the measured data to a series R/C model. This relatively high capacitance allows for off-the-shelf capacitors (>\textasciitilde 1pF) to be used for capacitive division and hence to facilitate measuring high voltages in the order of tens to hundreds of kVs. Fabricating the device on an insulating substrate would also enable capacitive division. For example, a quartz substrate ($\epsilon_r$=4.5 \cite{krupka1999complex}) with a typical thickness of 675$\mu m$ can be used to fabricate the sensor and form a capacitive divider on the same die; the quartz capacitance density (0.06$pF/mm^2$) is much lower than the AlN film capacitance density (280$pF/mm^2$) and allows approximately 5000:1 capacitive division for the same size sensor and capacitor. With a breakdown voltage of $100V$, this could enable voltage sensing up to $350kV_{rms}$, but one can choose to operate the sensor with voltages < $350kV_{rms}$ to extend the sensor operational lifetime.


\textbf {The low-cost OVS system can be built using our device and inexpensive optical components.} To our knowledge, we present the first OVS that uses an LED for operation, rather than a specialized light source such as an amplified
spontaneous emission source (ASE) \cite{gonccalves2019temperature, yang2017optical} or superluminescent LED (SLED) \cite{dante2016temperature} (see Table \ref{table:comparison}). Our low $Q$-factor device enables using a conventional broadband light source LED without the need of an optical filter, which allows sensor operation at higher powers near baseband with lower noise floors. This is because LEDs do not suffer from the same low-frequency excess noise that other narrowband light sources, such as lasers, do and can get very close to the shot noise limit \cite{rumyantsev2004low}. Additionally, the sensor can be interrogated at an angle or in a transmissive configuration; this would allow input and output fibers to be coupled directly to the sensor, without the necessity of using optical circulators or beamsplitters. Since this device is a relatively large size (mm-scale) it does not require precise alignment. Taken together, these properties can enable building an optical voltage sensing system based on our sensor and inexpensive optical components (LEDs and optical fibers), which could bring the cost of production below that of a commonly used instrument transformer or prior OVS systems \cite{gonccalves2019temperature, yang2017optical, dante2016temperature}.

\textbf {The OVS trades off \textit{energy per quanta} ($E_Q$) for temperature insensitivity and linearity.} Our sensor represents one extreme within the spectrum of all possible OVSs by deliberately excluding mirrors to reduce the $Q$-factor; the low $Q$-factor allows for our sensor to achieve better total harmonic distortion (THD) and temperature-induced relative  gain error ($\Delta \beta / \Delta T$) (without any compensation) than that of prior work, as shown in Table \ref{table:comparison}.

As we have shown, the noise efficiency of OVSs, represented by $E_Q$, can be traded off directly with nonlinearity and temperature dependence of the sensor output; $E_Q$ can be improved at the expense of an increase in nonlinearity and temperature sensitivity by increasing the sensor's $Q$-factor (supplementary materials). For a sensor that is desired to be operated with a low $E_Q$, the sensor design can be modified to incorporate mirrors on either side of the piezoelectric AlN thin film; this will improve the sensor $Q$-factor and hence improve the $E_Q$. Alternatively, the sensor thickness can be increased to increase the $Q$-factor (supplementary materials) and reduce $E_Q$, allowing the sensor to operate at higher input voltages.

\textbf{Our OVS operates within 6.6dB of the shot noise limit with no source feedback.} Previous systems, as shown by their high $E_Q$ values (see table \ref{table:comparison}), are operating well above the shot noise limit, wasting input photons. Our system operates within 6.6dB of the shot noise limit, with the primary excess noise due to the optical source. This efficient use of photons allows a low noise floor to be achieved despite a low modulation depth; this eliminates the need for closed-loop feedback to reduce the noise of the optical source, further reducing the cost and complexity of a potential OVS system.

\textbf {Limitations of the energy per quanta ($E_Q$) metric.} The $E_Q$ is a useful figure of merit when trying to use optical systems as sensors rather than (digital) communication devices because it is not a direct measure of energy per information (bit) used to express the energy efficiency in digital communication systems. Here, it represents a lower limit of noise performance for a shot-noise limited optical voltage sensor (OVS) operated at a fixed bandwidth ($BW$) and incident light power ($P_{in}$) and reveals the trade-offs between sensor SNR, $BW$, and $P_{in}$. In order to design an efficient OVS, the operating BW can be traded off directly for SNR, and $P_{in}$ can be traded for either $BW$ or SNR.




\begin{center}
\begin{table}
\caption{ Comparison of state-of-the-art optical voltage sensors. }
\label{table:comparison}
\begin{tabular}{c c c c c}

\hline & \cite{yang2017optical} & \cite{gonccalves2019temperature} & \cite{dante2016temperature} & \textbf{This work} \\ \hline
Architecture & Dual FBG & FBG & FBG & \textbf{Thin-film} \\ \hline
Light Source & Broadband ASE & Broadband ASE & SLED & \textbf{LED} \\ \hline 
$P_{in}$ & \textasciitilde 1mW  & 25mW$^c$ & $500\mu W^{f}$ & $\mathbf{110 \mu W}$ \\ \hline
$BW$ & 20kHz & 5kHz & 1kHz & $\mathbf{5kHz}$ \\ \hline
$SNR_{max}$ & 21dB & 3dB$^{d}$ & 54dB &\textbf{35dB}\\ \hline
$E_{Q}$ & \textasciitilde 370 & 2.5$\mu J$ & 1.9$pJ$ & \textbf{13}$\mathbf{pJ}$\\ \hline
THD & -47dB$^{a}$ & -23dB & -51dB & \textbf{-67dB} \\ \hline
$\frac{(\Delta \beta / \beta)}{\Delta T}$ & $44\%/^{\circ}C ^{b}$ & $0.2\%/^{\circ}C^{e}$ & $0.2\%/^{\circ}C^{e}$ & $\mathbf{0.011\%/^{\circ}C}$ \\ \hline
Compensation & FBG & Thermal screws & Bias point tracking & \textbf{None} \\ \hline

\end{tabular}
\\
\small \\ $^{a}$ Estimated from $V_{out}/V_{in}$ transfer function.  \\
\small $^{b}$ Maximum differential change in gain $\beta$ from $6^{\circ}C$ to $8^{\circ}C$. \\
\small$^{c}$ Input power not specified. Assume maximum power of typical ASE light source (500mW) with an insertion loss of 13dB, identical to our own. \\
\small $^{d}$ Estimated from PSD noise of 0.01V in 16Hz bin and signal (including spectral leakage) of 0.93V across 4 bins.\\
\small $^{e}$ Estimated from the temperature coefficient of PZT \cite{li2009determination}, which was not accounted for by the authors. \\
\small $^{f}$ Input power not specified. Assume power of typical SLED light source (10mW) with an insertion loss of 13dB, identical to our own. \\
\small FBG: Fiber Bragg Grating \\
$\Delta \beta / \beta$: Normalized sensor gain 
\end{table}
\end{center}


\section{Conclusion}

This work presents an AlN thin film based optical voltage sensor for power grid applications; the sensor is fabricated using standard microfabrication techniques. We demonstrate the advantages of this sensor in terms of nonlinearity and robustness to temperature variations, and articulate a figure of merit - the \textit{energy per quanta} ($E_Q$). The $E_Q$ fully captures the  trade-offs between sensor parameters, enabling the design of high-performance optical voltage sensors.

\section{Backmatter}

\begin{backmatter}
\bmsection{Funding}
Content in the funding section will be generated entirely from details submitted to Prism. Authors may add placeholder text in the manuscript to assess length, but any text added to this section in the manuscript will be replaced during production and will display official funder names along with any grant numbers provided. If additional details about a funder are required, they may be added to the Acknowledgments, even if this duplicates information in the funding section. See the example below in Acknowledgements.

\bmsection{Acknowledgments}
We would like to thank the staff of the Marvell Nanofabrication facility for supporting this work. This work was supported by the Hertz Foundation, the Berkeley Sensors and Actuators Center (BSAC), and the Chan-Zuckerberg Biohub. We would also like to thank Cem Yalçin and Ryan Kaveh for their conversations on mixed-signal circuits, and Ryan Rivers for his extensive support on design and fabrication. Michel Maharbiz is a Chan Zuckerberg Investigator.

\bmsection{Disclosures}

\medskip

\noindent The authors declare no conflicts of interest.

\bmsection{Data Availability Statement}
Data underlying the results presented in this paper are publicly available at \cite{edmunds2021data}.

\bmsection{Supplemental document}
See Supplement 1 for supporting content. 

\end{backmatter}


\bibliography{sample}
\end{document}


\maketitle

\section{Breakdown Field}

\begin{figure}[h]
  \includegraphics[width=4.5in]{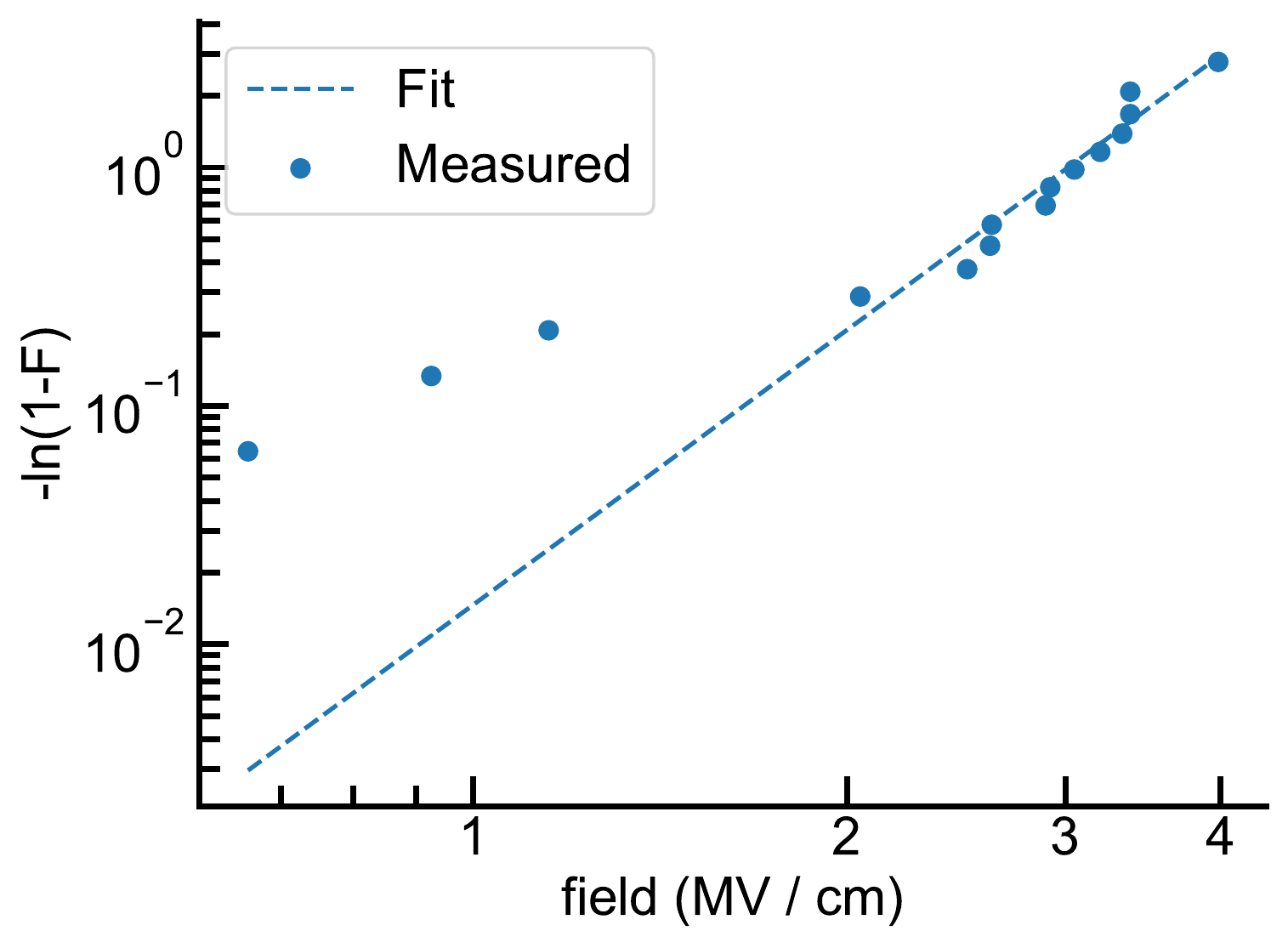}
  \caption{Cumulative number of failures vs. applied electric field (Weibull plot). F: CDF of the failure probability distribution. Raw data available in Dataset 5.}
  \label{fig:ovs-device-breakdown-field}
\end{figure}

The overall Weibull slope $\gamma$ for the breakdown field was measured to be 3.8, with a characteristic field $E_0$ of $3.0MV/cm$.

\section{Breakdown Charge}
\begin{figure}[h]
   \includegraphics[width=4.5in]{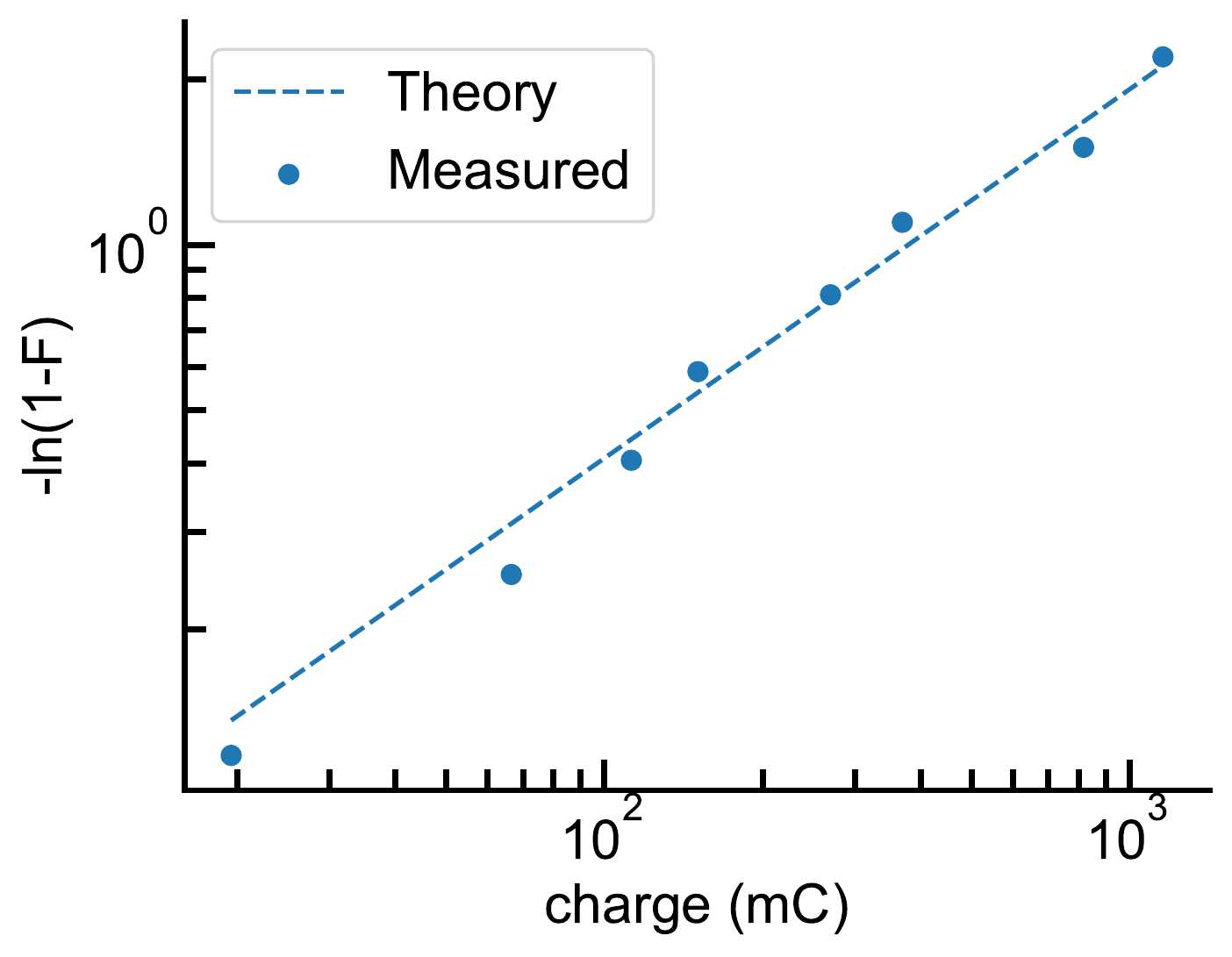}
  \caption{Cumulative number of failures vs. applied charge (Weibull plot). F: CDF of the failure probability distribution. Raw data available in Dataset 6.}
  \label{fig:ovs-device-breakdown-charge}
\end{figure}

The overall Weibull slope $\gamma$ for the breakdown charge was measured to be 0.67 with a characteristic field $E_0$ of $380mC$, which corresponds to a characteristic breakdown charge density of $12C/cm^2$. 

\section{Leakage current over voltage, temperature}
The leakage current reported in the device lifetime figure was at a temperature of 20C. To ensure that these conclusions are valid at elevated temperatures, and to predict lifetime versus operating condition, we measured the IV curve of one low-leakage devices at temperatures from 20C-100C. Above approximately 20V, the leakage current increases by an order of magnitude from 20C - 100C, but only by a factor of 2 between 20C and 60C. Colder temperatures would have lower leakage currents than shown here, and we may expect the average temperature over the device lifetime to be near room temperature.

\begin{figure}[h]
   \includegraphics[width=4.5in]{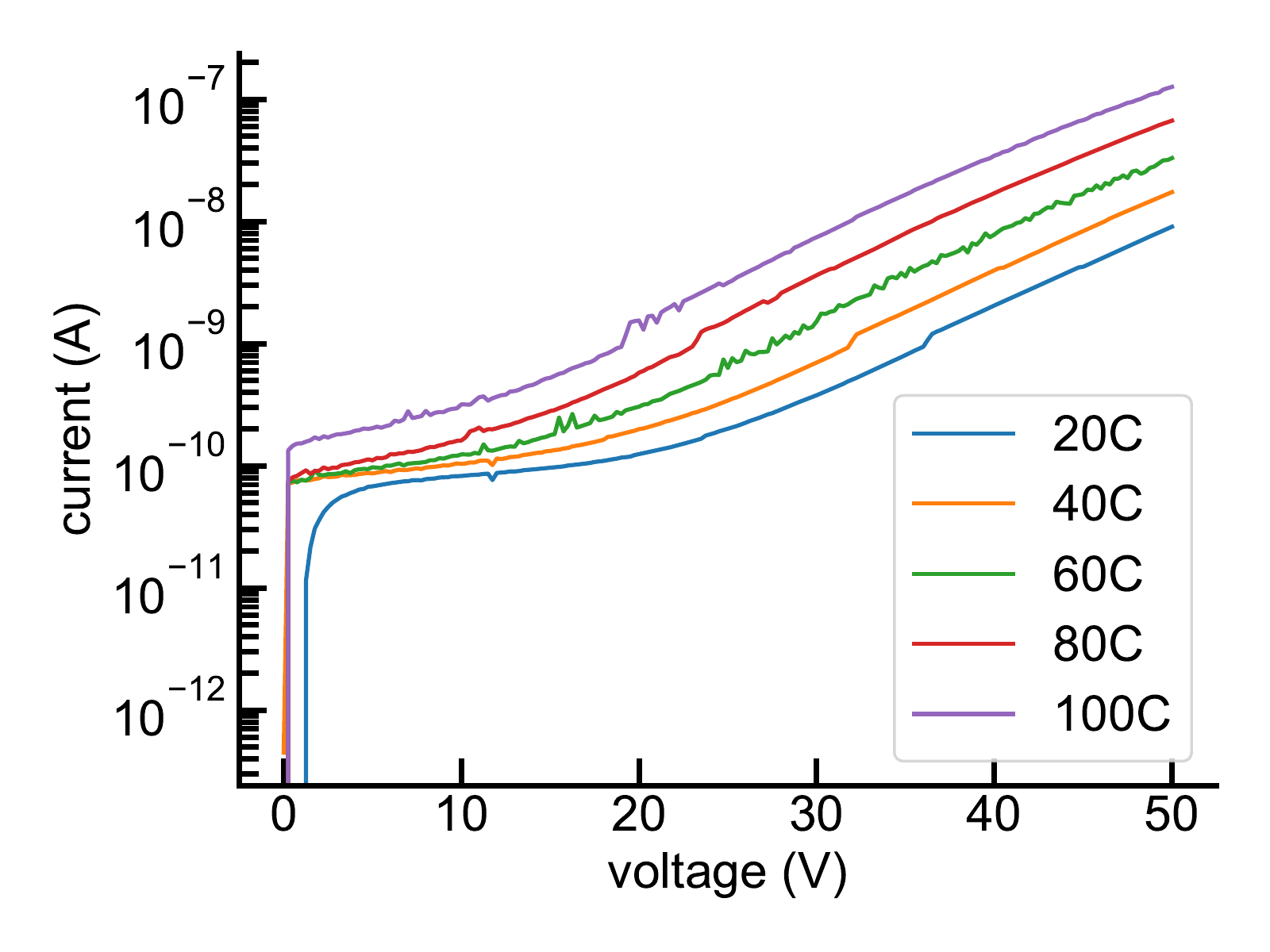}
  \caption{Leakage current vs. voltage at temperatures from 20C to 100C. Raw data available in Dataset 7.}
  \label{fig:ovs-device-breakdown-charge}
\end{figure}

\section{Device Impedance}
\begin{figure}
  \includegraphics[width=4.5in]{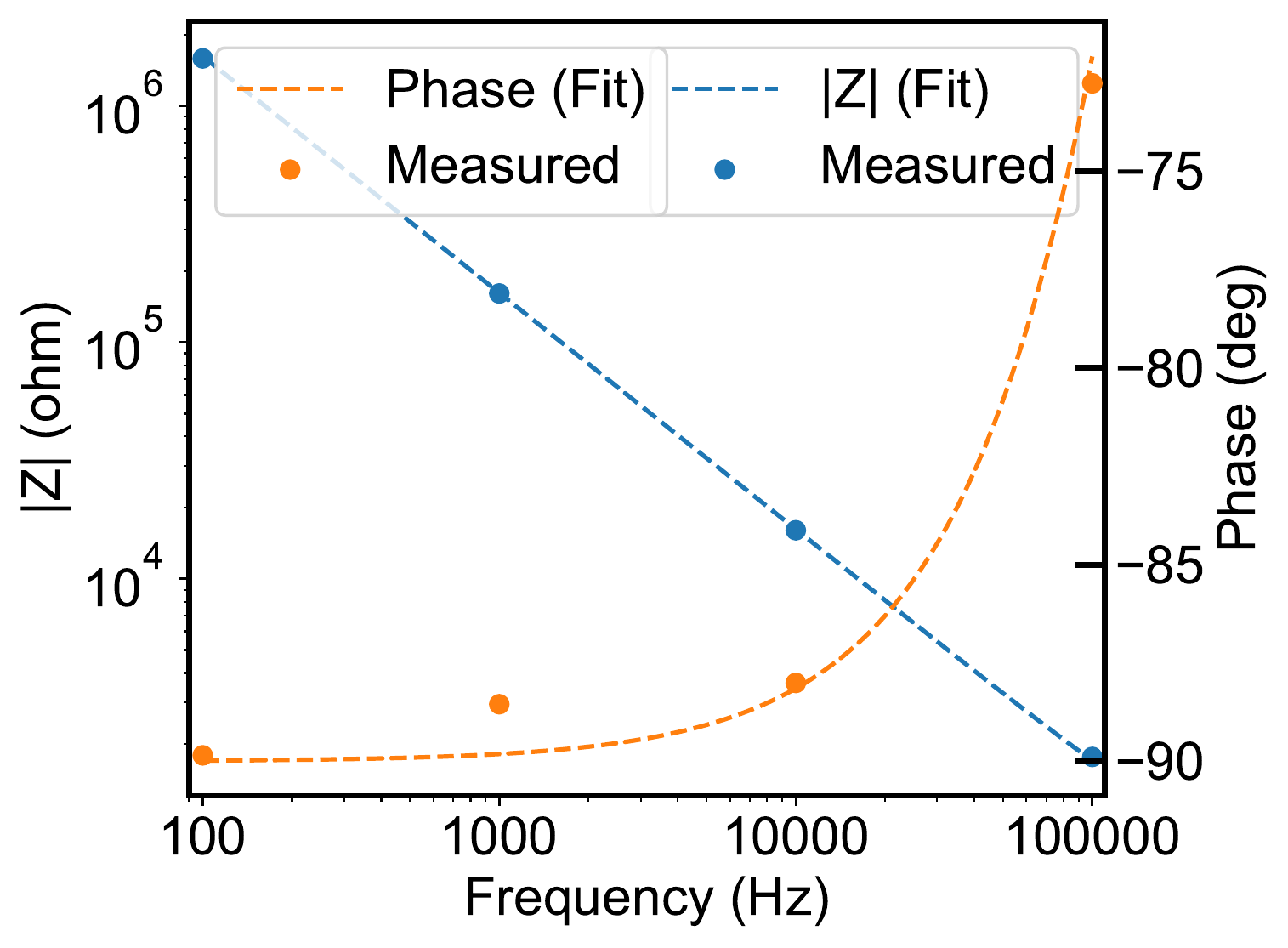}
  \caption{Representative device impedance vs frequency from 10Hz-500kHz. Theoretical curve was fit using an R/C series model, fitting to the logarithm of the magnitude of Z and the phase directly using the python symfit package. Fitted series resistance was 389Ohms and fitted capacitance was 997pF. Raw data in Dataset 8.}
  \label{fig:ovs-device-impedance}
\end{figure}

Device impedance was measured using an Agilent 4284A LCR meter, and the measured impedance of a typical device, along with the best-fit RC model is shown in Fig. \ref{fig:ovs-device-impedance}.

\section{Device Reflectance Spectra, Model Fitting}
\begin{figure}
  \includegraphics[width=4.5in]{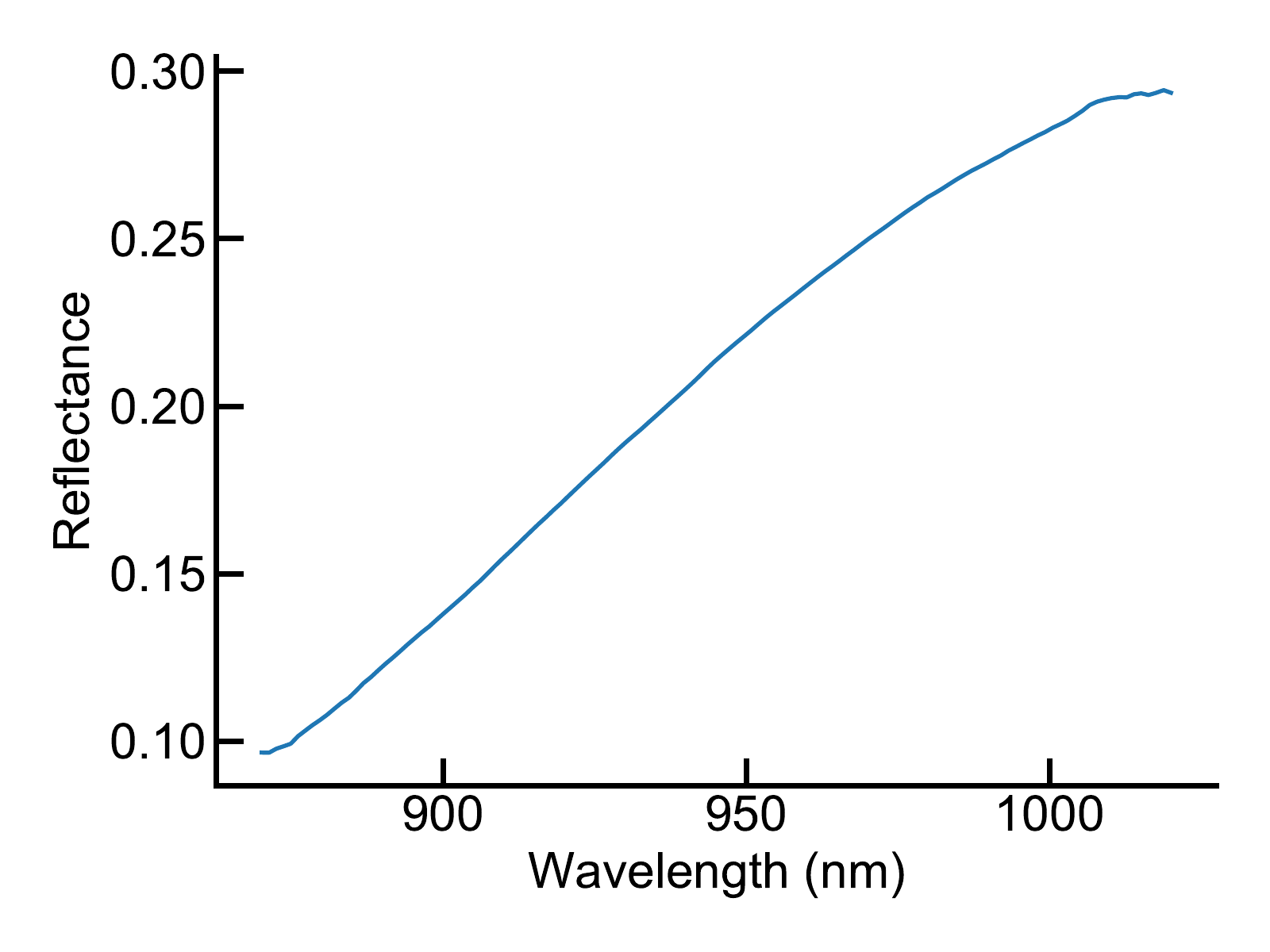}
  \caption{Measured reflectance spectra of optical modulator. Raw data available in Dataset 9.}
  \label{fig:ovs-reflection-spectra}
\end{figure}

The reflectance of a single device was measured in the range 870-1020nm, and the reflectance is shown in Fig. \ref{fig:ovs-reflection-spectra}. This was fit to a Lorentzian function with Q and $R_{max}$ as fitting parameters which are shown in table \ref{table:parameters}. 

\section{Fabrication Process}

A detailed fabrication process is given below with settings used, using the same numbering as Fig. 2 in the main text. \\
 
Prior to this process, we deposited 0.3um of UV210 resist, and patterned  alignment marks for layer-to-layer alignment in the ASML 5500/300 DUV stepper. We then etched 6 alignment marks into the Si substrate in a Poly-Si etcher (TCP 9400SE Lam Research) with 300W plasma power, 150W substrate power, 50 standard cubic centimeters per minute (sccm) chlorine gas ($Cl_2$), 150sccm hydrogen bromide (HBr), 4sccm helium (He), at a process pressure of 12mTorr for 40 seconds, with a target etch depth of 120nm.

\begin{enumerate}
\item
  Al Sputtering and Anneal- We first sputter etched the backside of the wafer at 200W RF Power / 10sccm argon (Ar) flow for 120 seconds. Without breaking vacuum, we then sputtered \textasciitilde 300nm of aluminum (Al) at 4.5kW DC power / 15sccm Ar flow for 60 seconds. Target was conical and 99.999\% pure. To make the backside Al contact ohmic to minimize contact resistance and nonlinearity, we annealed the wafer in an AccuThermo 610 at 300$^{\circ}$C for 15 minutes, with an initial ramp rate of \textasciitilde 20$^{\circ}$C/s at atmospheric pressure flowing pure $N_2$. The specific contact resistance at 0V dropped from $600 M\Omega \cdot \mu m^2$ to $5 M\Omega \cdot \mu m^2$ (\textasciitilde 110x improvement), and became linear instead of rectifying. 
\item
  AlN Deposition - Prior to each deposition, we ran a bare Si conditioning wafer Al sputter to remove built-up AlN from the target. We sputtered the target at 4kW / 6sccm Ar flow for 5 minutes, followed by a brief (30 second) period where we introduced $N_2$ gas, flowing 7 sccm Ar and 22 sccm $N_2$ at 4kW power. We then sputter etched the frontside of the device wafer at 200W RF Power / 10sccm Ar flow for 120 seconds. Without breaking vacuum, we then sputtered \textasciitilde 300nm of aluminum nitride (AlN) at 4.5kW, flowing 7sccm Ar and 22 sccm $N_2$ for 420 seconds at a chamber pressure of 5mTorr. The chamber used was dedicated to only deposit AlN.
\item
  Topside ITO Deposition - We deposited ITO from an 8" target at 500V DC and 2mTorr, flowing 40sccm of Ar into the chamber for 60 seconds, for a target ITO thickness of 20nm. 
\item
  Topside ITO patterning - We spun on 0.3um of UV-210 photoresist in an automated Picotrack system, exposed a full wafer of 10mm x 10mm dies in an ASML DUV stepper (Model 5500/300) at $20mJ/cm^2$ dose. We then developed the resist in MF26A developer for 60 seconds in the same automated Picotrack coating system, and UV-hard baked the resist in a Fusion Systems M200PCU. We then ion milled the ITO layer in a Pi Scientific ion mill at 500V / 250mA at an angle of 20 degrees away from normal incidence for 60 seconds, followed by an etch at 70 degrees away from normal incidence for 20 seconds. The resist was then removed by pressurized and heated NMP (20MPa / 80C / 1000rpm). 
\item
  Al bond pad patterning - We then spun on 1um of LoR-5A liftoff resist in an automated track coater (SVG 8626) for subsequent liftoff, followed by 0.87um of UV210 resist in an automated resist coating system (Picotrack). We then patterned the bond pads and traces connecting the pads to the ITO layer in an ASML 5500/300 DUV stepper. We then evaporated 20nm titanium (Ti) / 300nm Al in an electron beam evaporator (CHA Solutions). We deposited the Ti at 10kV beam voltage / 90mA beam current and the Al at 10kV / 50mA. We monitored the thickness with a 6MHz crystal monitor.
\end{enumerate} 

Following the main fabrication process, individual 10mm x 10mm chips were coated with 2um of protective i-Line resist and then diced in an automated dicing saw (Disco DAD324). The resist was then removed with acetone, and individual chips removed for subsequent bonding onto a PCB. We bonded the Al-coated backside of each die to an IPA-cleaned gold (Au) 10mm x 10mm pad on a custom PCB with two-component silver epoxy (MG Chemicals 8331), cured at 50$^{\circ}$C for 30 minutes. We then wirebonded the top 150um x 150um Al bond pad to a bond pad on the PCB with 50$\mu m$ Al wire with a manual wirebonder (West Bond Model 7400B). 

\section{Device Transfer Function, Gain}
This device operates by a shift the resonant frequency, and the change in output reflectance $\Delta R$ can be linearly related to the input voltage $V_{in}$ through the gain $\beta$: ($\Delta R = \beta \cdot V_{in}$) . We can approximate the lineshape as a Lorentzian (this approximation is equivalent to a small-angle approximation, and gets better with increasing $Q$):

\begin{equation}
    R(x) \approx R_{max}\left(1 - \frac{1}{1+ x^2}\right)
\end{equation}

Where x, the normalized frequency, is equal to $\frac{\omega_r - \omega_{in}}{FWHM/2}$, where $FWHM$ is the full-width-half-max of the spectral lineshape, $\omega_r$ is the resonant frequency at which destructive interference occurs, and $\omega_{in}$ is the input frequency. By expanding this in terms of derivatives and finite differences, we find:

\begin{equation}
    \Delta R \approx \frac{dR}{dx} \cdot \Delta x
\end{equation}

\begin{equation}
    \frac{dR}{dx} = \frac{-2 R_{max} x}{\left( 1 + x^2 \right)^2}
\end{equation}

\begin{equation}
    \left(\frac{dR}{dx}\right)_{max} = \mp \frac{-2 R_{max}3 \sqrt{3}}{8}
\end{equation}

\begin{equation}
    \Delta x = \frac{\Delta \omega_r}{FWHM/2}
\end{equation}

\begin{equation}
    \Delta \omega_r \approx \omega_{r0} \left( \frac{\Delta L}{L_0} + \frac{\Delta n}{n_0} \right)
\end{equation}

A change in physical length occurs due to the converse piezoelectric effect:

\begin{equation}
    \frac{\Delta L}{L_0} = \frac{d_{33} * V_{in}}{L_0}
\end{equation}

and a change in refractive index due to the Pockels effect:

\begin{equation}
    \frac{\Delta n}{n_0} = \frac{n_0^2 \cdot r_{33} * V_{in}}{2 L_0}
\end{equation}

Combbining these equations together at the point of maximum gain, and recognizing that $Q = \omega_{r0} / FWHM$, we find that:

\begin{equation}
    \beta = \frac{3 \sqrt{3}}{4} \frac{R_{max} Q}{L_0}\left(d_{33} + \frac{1}{2} n_0^2 r_{33} \right)
\end{equation}

\subsection{Q-factor Length dependence}
By solving the Fresnel equations \cite{hecht2002optics} for a cavity with equal mirror reflectivities, we can find an exact form of the cavity reflected intensity:

\begin{equation}
    R = 1 - \frac{1}{1 + F * sin^2(\phi)}
\end{equation}

Where $F$ is the coefficient of finesse, which can be written in terms of the mirror reflectivity $R_m$:

\begin{equation}
    F = \frac{4 R_m}{\left(1 - R_m\right)^2}
\end{equation}

We can solve the above equations for the full-width-half-max of the spectra, and can obtain a form of the quality factor which depends on the incident wavelength $\lambda_{in}$ and the length and refractive index of the cavity:

\begin{equation}
    Q = \frac{2\pi n L}{FWHM_{\phi} \lambda_{in}}
\end{equation}

Where $FWHM_{\phi} = 2 Arcsin\left(\frac{1}{\sqrt{2 + F}} \right)$. 
 
\begin{table}
\begin{center}
\caption{Parameters used in the calculation of sensor gain for analytical model and Fresnel model}
\label{table:parameters}
\begin{tabular}{c c}

Parameter & Value \\ \hline
$r_{33}$ & $1pm/V$ \cite{graupner1992electro} \\ \hline
$d_{33}$ & $5pm/V$ \cite{lueng2000piezoelectric} \\ \hline
$L_0$ & $295nm^a$ \\ \hline
$n_0$ & 2.13 \cite{pastrvnak1966refraction} \\ \hline
$R_{max}$ & $0.39^b$ \\ \hline
$Q$ & $4.0^b$ \\ \hline

\end{tabular} \\

\small $^{a}$ Measured using interferometry. \\
\small $^{b}$ Best fit of Lorentzian to measured reflection spectra. \\
\end{center}
\end{table}

\section{Gain Nonlinearity}
To quantify the nonlinearity, we assume that the device is biased at the wavelength which corresponds to maximum linearity. This corresponds to the best-case device nonlinearity, and accounts only for the nonlinearity in the reflectance lineshape. In practice, this will be dominant as the quality factor becomes large. At very low Q-factors, other sources of nonlinearity, such as material nonlinearity, may be dominant. 

In terms of normalized frequency / wavelength $x_0$ this is $\pm \frac{1}{\sqrt{3}}$. The nonlinearity can be quantified in terms of gain compression, or normalized change in linear gain:
 
\begin{equation}
    \label{eq:dGnorm}
    \frac{\Delta \beta}{\beta_{max}} = -\frac{9}{4} \Delta x^2
\end{equation}

where $\Delta H$ is the absolute deviation in gain and $H_{max}$ is the gain at maximum linearity, which is also equal to the maximum gain. $H_{max}$ is $\mp \frac{3\sqrt{3}}{8}$. We can now combine this with the energy per quanta, defined in the main text in terms of the modulation depth $\Delta R$. Without loss of generality, we can write $\Delta R \approx G(x_0) * \Delta x$. A change in voltage will change the normalized frequency, which results in a change in reflectance due to shifting of the reflectance spectrum. Combining the terms in the energy per quanta $E_Q = \frac{2q \cdot R_0}{\Re \Delta R^2} = \frac{2q \cdot R_0}{\Re G(x_0)^2*\Delta x^2}$ and the above expression we have:

\begin{equation}
    \label{eq:nl-es-tradeoff}
    E_Q * \frac{\Delta \beta}{\beta_{max}} \approx \left(\frac{2q \cdot R_0}{\Re H_{max} ^2*\Delta x^2}\right) * \left(-\frac{9}{4} \Delta x^2 \right) = 4 \sqrt{3} \frac{q \cdot R_0}{\Re}
\end{equation}

The product of the energy per quanta and the nonlinearity is equal to a constant. In other words $E_Q = const * \frac{1}{NL}$. A better (lower) energy per quanta can only be achieved at the cost of a worse (higher) nonlinearity.

If the incident wavelength is not at the optimal wavelength, as will usually be the case, this result is worse by a factor of approximately $2 \frac{x_{os}}{\Delta x}$, where $x_{os}$ is the normalized offset frequency. Alternatively, we can say that $E_Q$ trades off against the \textit{square} of the nonlinearity, as quantified by the relative change in gain.

\section{Gain temperature dependence}
The prior section generalizes to anything which changes the resonant frequency of the device - it need not be voltage. We can thus take the results above to apply for temperature as well - the energy per quanta trades off linearly temperature sensitivity, as quantified by the normalized change in the gain $E_Q \propto 1 / \frac{\Delta G}{G_{max}}(T)$.

By expanding the relative gain change $\frac{dH}{H}$ in terms of the resonant frequency:

\begin{equation}
    \frac{d\beta}{\beta} = -\frac{9}{4}\Delta x^2
\end{equation}

Where $\Delta x$ is the shift in normalized resonant frequency away from its optimal value, $dH$ is the absolute change in gain and $H$ is the gain.

\begin{equation}
    \Delta x = \frac{\Delta \omega_r}{\gamma /2}
\end{equation}

Where $\omega_r$ is the resonant frequency $(rad/s)$, and $\gamma$ is the full-width-half-max linewidth of the spectra.

\begin{equation}
    \Delta \omega_r = \omega_r \Delta T \left( \alpha_n + \alpha_L \right)
\end{equation}

Where $\Delta T$ is the change in temperature, $\alpha_n$ is the relative change in refractive index (sometimes called the thermo-optic coefficient), and $\alpha_L$ is the Putting these three equations together, and recalling that $Q \equiv \omega_r / \gamma$ we find the relative change in gain to be:

\begin{equation}
    \frac{d\beta}{\beta} = -9 Q^2 \Delta T^2 \left(\alpha_n + \alpha_L\right)^2
\end{equation}

For a the parameters used to model this device, we expect the change in gain over the temperature range used to be $0.05\%$




\bibliography{sample}